\documentclass[%
 reprint,
superscriptaddress,
 amsmath,amssymb,
 aps,prl
]{revtex4-2}

\usepackage{graphicx}
\usepackage{dcolumn}
\usepackage{bm}

\usepackage[colorlinks = true]{hyperref}
\usepackage{physics}

\usepackage[english]{babel}

\makeatletter
\def\bbl@set@language#1{%
  \edef\languagename{%
    \ifnum\escapechar=\expandafter`\string#1\@empty
    \else\string#1\@empty\fi}%
  \@ifundefined{babel@language@alias@\languagename}{}{%
    \edef\languagename{\@nameuse{babel@language@alias@\languagename}}%
  }%
  \select@language{\languagename}%
  \expandafter\ifx\csname date\languagename\endcsname\relax\else
    \if@filesw
      \protected@write\@auxout{}{\string\select@language{\languagename}}%
      \bbl@for\bbl@tempa\BabelContentsFiles{%
        \addtocontents{\bbl@tempa}{\xstring\select@language{\languagename}}}%
      \bbl@usehooks{write}{}%
    \fi
  \fi}
\newcommand{\DeclareLanguageAlias}[2]{%
  \global\@namedef{babel@language@alias@#1}{#2}%
}
\makeatother

\DeclareLanguageAlias{en}{english}

\begin{document}


\title{Optimizing performance of quantum operations with non-Markovian decoherence: \\ the tortoise or the hare?}
\author{Eoin P. Butler}
 \affiliation{School of Physics, Trinity College Dublin, College Green, Dublin 2, Ireland}
 \affiliation{Trinity Quantum Alliance, Unit 16, Trinity Technology and Enterprise Centre, Pearse Street, Dublin 2, Ireland}
\author{Gerald E. Fux}%
\affiliation{SUPA, School of Physics and Astronomy, University of St Andrews, St Andrews KY16 9SS, United Kingdom}%
\affiliation{Abdus Salam International Center for Theoretical Physics (ICTP), Strada Costiera 11, 34151 Trieste, Italy}%
\author{Carlos Ortega-Taberner}
 \affiliation{School of Physics, Trinity College Dublin, College Green, Dublin 2, Ireland}
 \affiliation{Trinity Quantum Alliance, Unit 16, Trinity Technology and Enterprise Centre, Pearse Street, Dublin 2, Ireland}
\author{Brendon W. Lovett}
\affiliation{SUPA, School of Physics and Astronomy, University of St Andrews, St Andrews KY16 9SS, United Kingdom}
\author{Jonathan Keeling}
\affiliation{SUPA, School of Physics and Astronomy, University of St Andrews, St Andrews KY16 9SS, United Kingdom}%
\author{Paul R. Eastham}
 \affiliation{School of Physics, Trinity College Dublin, College Green, Dublin 2, Ireland}
 \affiliation{Trinity Quantum Alliance, Unit 16, Trinity Technology and Enterprise Centre, Pearse Street, Dublin 2, Ireland}
\date{\today}
\begin{abstract}
The interaction between a quantum system and its environment limits our ability to control it and perform quantum operations on it. We present an efficient method to find optimal controls for quantum systems coupled to non-Markovian environments, by using the process tensor to compute the gradient of an objective function. 
We consider state transfer for a driven two-level system coupled to a bosonic environment, and  characterize performance in terms of speed and fidelity. This allows us to determine the best achievable fidelity as a function of process duration. 
We show there can be a trade-off between speed and fidelity, and that slower processes can have higher fidelity by exploiting non-Markovian effects.
\end{abstract}

\maketitle

The control of quantum systems using time-dependent Hamiltonians is crucial for quantum technologies~\cite{glaser_training_2015}, enabling the implementation of state transfer and gate operations. An important task is to establish how optimal performance can be achieved for such processes. In an ideal closed quantum system perfect operations are possible given sufficient time~\cite{deffner_quantum_2017}. A speed limit arises because physical Hamiltonians are bounded, so that energy-time uncertainty gives a maximum rate of time-evolution and hence a minimum operation time. Beyond this ideal case, however, other considerations arise. One is a desire for reliable operation when precise control cannot be guaranteed; this can be achieved by using robust control techniques~\cite{propson_robust_2022} or adiabatic processes~\cite{benseny_adiabatic_2021,Torrontegui2013:Review}. Another is the impact of decoherence and dissipation. In the standard Markovian approximation such processes give a cumulative loss of information with time. Thus, one generally expects fast operation to be desirable to minimize information loss, although there are notable exceptions, where slower operation allows access to a decoherence-free subspace~\cite{basilewitsch_optimally_2020}. In the present Letter, we show that fast operation is not always desirable in non-Markovian systems, because slower operation can enable information backflow to be harnessed to increase fidelity. 

 To provide a concrete demonstration of the trade-off between speed and fidelity in non-Markovian systems we use numerical optimal control to explore the achievable performance  for a system consisting of a driven qubit interacting with a bosonic environment. Optimal control~\cite{koch_quantum_2022} involves determining a set of time-dependent control fields that maximize an objective function such as the fidelity. Here, we show that this can be done effectively in a non-Markovian system, using an extension of our previously introduced process tensor approach~\cite{fux_efficient_2021} to efficiently compute the gradient of the objective function. This allows us to repeatedly optimize over many hundreds of control parameters for different process durations, and so find the achievable fidelity as a function of the process duration. Our results reveal a marked improvement in fidelity for process durations $T$ longer than a value $T_0$, set by the speed-limit in the closed system~\cite{deffner_quantum_2017}. We further compute a widely-used measure of non-Markovianity, based on trace-distance~\cite{laine_measure_2010}, and show that the improved fidelity coincides with an increase in non-Markovianity. Since many types of device exhibit regimes of non-Markovian decoherence our results could enable performance improvements across a range of quantum technologies~\cite{acin_quantum_2018}, including superconducting circuits~\cite{fischer_time-optimal_2019,goerz_charting_2017}, spins~\cite{dolde_high-fidelity_2014}, quantum dots~\cite{fux_efficient_2021,wilson-rae_quantum_2002}, color centers in diamond~\cite{norambuena_quantifying_2020}, and cold atoms~\cite{choi_optimal_2014,bason_high-fidelity_2012,treutlein_microwave_2006}. 

Our approach applies to an open quantum system comprising a few-state system Hamiltonian $H_S$ interacting with an environment (bath) with Hamiltonian $H_B$ through a coupling $H_{SB}$. Models of this form are typically treated under the assumptions of weak system-bath coupling and short bath correlation time (Born-Markov). This implies that information about the system only flows away into the environment, and does not return,  leading to simple time-local evolution equations for the system reduced density matrix $\rho_S$~\cite{breuer_theory_2007}.  To avoid making these approximations we use process tensors (PT), which are general objects that encode the complete influence of the environment. Our approach applies to any open quantum system for which the PT can be computed as a matrix-product-operator (PT-MPO) with low bond dimension.  

Time-local evolution equations for $\rho_S$ have been used to explore performance limits~\cite{caneva_optimal_2009,del_campo_quantum_2013,deffner_quantum_2013} using optimal control~\cite{koch_controlling_2016} in various problems. Studying optimal control beyond their applicability has so far been difficult. One approach is to expand the system to incorporate a few modes of the environment, which are then treated exactly, with the remaining environment modes providing Markovian damping. This approach has been used to study controllability~\cite{reich_exploiting_2015}, optimal control~\cite{rebentrost_optimal_2009,floether_robust_2012} and speed limits~\cite{deffner_quantum_2013}. However, it becomes intractable when one considers more than a few modes of the environment. Another method involves computing the time-local dissipator describing non-Markovian dynamics using lowest-order perturbation theory~\cite{hwang_optimal_2012} or the hierarchical equations-of-motion~\cite{mangaud_non-markovianity_2018}. These techniques have been used within optimal control to demonstrate fidelity increases, and can be effective when the environment spectral density can be approximated by a small number of Lorentzians. Another type of approach~\cite{schmidt_optimal_2011} uses the stochastic Liouville equation, but is limited by the need to average over a large and \emph{a priori} unknown number of trajectories. Some works~\cite{ohtsuki_non-markovian_2003,ying-hua_optimal_2016,cui_optimal_2008} have obtained optimal protocols under the assumption that the dissipation is described by a fixed time-local dissipator, such as that for a pure dephasing channel~\cite{chin_quantum_2012,bylicka_non-markovianity_2014}. An issue, however, is that for optimal control one must consider the effect of the time-dependent control fields on the dissipative processes~\cite{addis_problem_2016}.

\begin{figure}
\includegraphics[width=\linewidth]{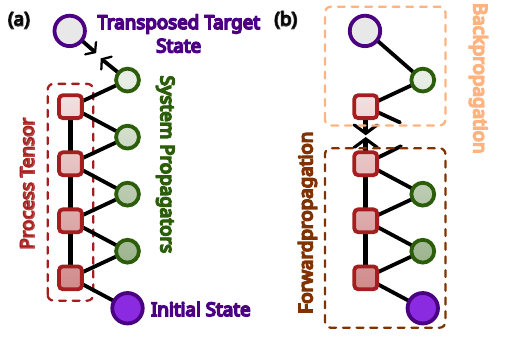}
\caption{\label{fig:ptdiagrams} (a) Tensor network representation of the propagation of an initial state over four time steps and calculation of the fidelity. Time increases going up the diagram. The evolved density matrix is given by this diagram without the target state node. The fidelity is obtained by joining the legs indicated by arrows. (b) The derivative of the fidelity with respect to the penultimate propagator is given by joining the legs indicated by the arrows, contracting a pair of indices in the product of one tensor formed during a forward propagation with one formed during a backward propagation.}
\end{figure}

To overcome these challenges, we extend the process tensor method described in our previous work~\cite{fux_efficient_2021} for solving the dynamics of non-Markovian open systems. The process tensor (PT)~\cite{pollock_non-markovian_2018,supplement} is a multilinear map from the set of all possible system control operation sequences to the resulting output states, constructed by discretizing the time evolution into a series of time steps. It can often be computed in matrix-product operator form with low bond dimension using algorithms which systematically discard irrelevant correlations~\cite{verstraete_matrix_2004,orus_practical_2014}. For a Gaussian bosonic environment efficient representations can be found using the methods introduced in~\cite{makri_tensor_1995,makri_tensor_1995-1, strathearn_efficient_2018, jorgensen_exploiting_2019,fux_efficient_2021}, which are often effective for smooth spectral densities. A range of methods for bosonic, fermionic, and spin environments are now available~\cite{banuls_matrix_2009, strathearn_modelling_2020, lerose_influence_2020, cygorek_simulation_2022, sonner_influence_2021, ye_constructing_2021, white_non-markovian_2021, thoenniss_non-equilibrium_2022, OQuPy2022, thoenniss_efficient_2022, ng_real_2022, link_open_2023}. Once computed, the PT can be contracted to time-evolve $\rho_{S}$ for any $H_{S}$, as shown in Fig.~\ref{fig:ptdiagrams}(a). In this tensor network diagram~\cite{orus_practical_2014} each node represents a tensor, each leg represents an index, and connections between legs correspond to contractions. The diagram is in Liouville space so that operators such as $\rho_S$ are rank-1 objects (i.e. objects with a single index), and maps between operators, such as the propagators across each time step $U_t=e^{\Delta t \mathcal{L}_S(t)}$ with $\mathcal{L}_S(t)\bullet=-i[H_S(t),\bullet]$,  are rank-2. The PT is the region in the dashed box, which can be contracted with an initial density matrix and the Trotterized propagators~\cite{trotter_product_1959}
to obtain $\rho_S$ at later times. 

For optimal control we must define an objective function. In the following we consider the example of a state transfer process, for which we use the fidelity $\mathcal{F}$~\cite{benenti_principles_2018} between the desired target state and the obtained final state $\rho_f=\rho_S(t_f)$. To simplify the presentation of the method we assume a pure target state, with density matrix $\sigma=|\sigma\rangle\langle\sigma|$, in which case $\mathcal{F}=\langle \sigma|\rho_f|\sigma\rangle$ is a linear function of the final state. In Liouville space it is the scalar product of the vectors representing the final density matrix and the transposed target density matrix, corresponding to the diagram in Fig.~\ref{fig:ptdiagrams}(a). The generalization of the method to nonlinear objective functions, such as the fidelity for a mixed target state, is straightforward~\cite{supplement}. 

An optimal protocol is found by numerically maximizing $\mathcal{F}$ over $N$ control parameters which determine the system propagators at each discrete time step. Such numerical optimization becomes significantly faster if one can efficiently calculate the gradient of $\mathcal{F}$ with respect to the $N$-dimensional vector of control parameters, $\nabla\mathcal{F}$. A naive calculation of $\nabla\mathcal{F}$ requires of order $N$ evaluations of $\mathcal{F}$, strongly limiting the size of problems that can be treated. Such gradients can however be efficiently obtained using the adjoint method~\cite{plessix_review_2006}, which has been applied to unitary or Markovian dynamics~\cite{khaneja_optimal_2005,mishra_control_2021}. In the GRAPE algorithm~\cite{khaneja_optimal_2005}, for example, the gradient is computed by combining states stored during a forwards-in-time propagation from the initial state with those stored during a backwards-in-time propagation from the target state. Crucially, one may see that the tensor network representation of $\mathcal{F}$ in Fig.~\ref{fig:ptdiagrams}(a) leads immediately to the generalization required for non-Markovian dynamics. This diagram represents the fidelity as a multilinear functional of the propagators; thus the derivative with respect to the propagator at a given time step is the same diagram with the node for that propagator removed. As shown in Fig.~\ref{fig:ptdiagrams}(b), and discussed further in~\cite{supplement}, all such derivatives can be computed by combining rank-2 tensors stored as the network is evaluated forwards in time with those stored during a backward propagation. The key difference compared to the adjoint method for unitary or Markovian dynamics is that we propagate a rank-2 tensor rather than the state $\rho_S$. The additional index in this tensor encodes information about the past and future dynamics and enables optimization of non-Markovian evolutions. 

To illustrate the general principle, 
we consider an example of optimal state transfer in a driven two-level system. Our system Hamiltonian is $H_S=h_x(t)s_x+h_z(t)s_z$, where the fields $h_{x,z}(t)$ are the controls. We treat them as piecewise constant with values $h_{x,z}(t_n)$ at the time steps $t_n=n\Delta t$. We seek to determine the control fields which steer the state $\rho_S(t)$ from a prescribed initial state (the state with $s_x=-1$) to a target state (the state $s_x=+1$). 
The bath is a set of harmonic oscillators, and the coupling is $H_{SB}=s_z\sum_q(g_qb_q+g_q^*b_q^\dagger)$.
This model is known as the spin-boson model, and it describes many physical systems~\cite{leggett_dynamics_1987,de_vega_dynamics_2017}. For definiteness, we suppose it represents an optical transition on a quantum dot driven by a laser pulse~\cite{ramsay_review_2010}, in which case $h_z$ and $h_x$ are the time-dependent detuning and Rabi frequency respectively.  In the quantum dot the bosonic modes are acoustic phonon modes, for which we use the super-Ohmic spectral density  $J(\omega)=2\alpha\omega^3\omega_c^{-2}\exp\left[-\omega^2/\omega_c^2\right]$ with $\alpha=0.126$ and $\omega_c=3.04\,\text{ps}^{-1}$~\cite{ramsay_phonon-induced_2010,luker_influence_2012}, and use a bath temperature of 5\,K. 

\begin{figure}
\includegraphics{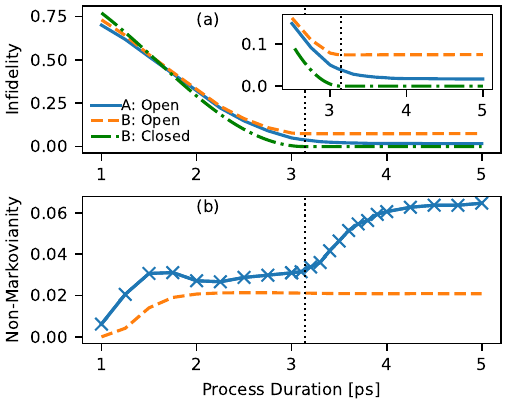}
\caption{\label{fig:fidelity}(a) Infidelity, $1-\mathcal{F}$, for state transfer as a function of process duration with bounded control fields. Results are shown for the optimal protocol of the non-Markovian spin-boson model (control A) and for driving with a constant field (control B). The latter is an optimal protocol in the closed system. Blue/solid: infidelity of control A in the spin-boson model. Orange/dashed: infidelity of control B in the spin-boson model. Green/dot-dashed: infidelity of control B in the closed system. (b) Non-Markovianity measure as a function of process duration in the spin-boson model for controls A (blue/solid) and B (orange/dashed). The vertical dotted lines are the speed limit time $T_0$ of the closed system.}
\end{figure}

We have determined optimal controls $h_x(t_n),h_z(t_n)$ by numerically maximizing $\mathcal{F}$ using the L-BFGS-B algorithm~\cite{fletcher_practical_2000,virtanen_scipy_2020}, with the gradient and fidelity computed with the PT. To explore the impact of speed on fidelity we perform this optimization for different process durations $T$, with bounds on the fields,
$|h_x|\leq h_x^{\mathrm{max}}=5\,\text{ps}^{-1}$ and $|h_z|\leq h_z^{\mathrm{max}}=1\,\text{ps}^{-1}$,
to consider the realistic case in which the Hamiltonian, and therefore the speed of the unitary evolution, is restricted.  We note that in the unitary case, the speed limit for state transfer has been identified from the rate of convergence of the Krotov algorithm~\cite{caneva_optimal_2009}. Here we consider a fully non-Markovian dissipative system, and use numerically converged optimization to identify the speed limit from the behavior of the fidelity against process duration. 
Figure \ref{fig:fidelity}(a) shows the resulting infidelity, $1-\mathcal{F}$ for this optimized protocol, denoted ``Control A''. 

It is interesting to compare the optimized results with those for the protocols which are optimal for unitary dynamics in the closed system~\cite{hegerfeldt_driving_2013}, denoted ``Control B'', which we used as the starting point for the optimization. These protocols can be understood by noting that for unitary dynamics one can achieve the state transfer with infidelity zero by setting $h_x$=0 and requiring an area $\pi$ for the time integral of the field $h_z(t)$. Such a protocol is  optimal if it satisfies the constraint $|h_z(t)|< h_z^{\mathrm{max}}$ for all times, which is possible when $T$ is greater than the speed limit time $T_0=\pi/ h_z^{\mathrm{max}}$ (which saturates both the Mandelstam-Tamm and Margolus-Levitin bounds~\cite{deffner_quantum_2017}). Among protocols satisfying the above condition, we choose $h_z(t)=\pi/T$. For $T<T_0$ the state transfer is not fully achievable and the optimal protocol is that with $h_z(t)=\pm h_z^{\mathrm{max}}$. Thus, in both regimes, we have an optimal protocol with constant fields. As can be seen in Fig.~\ref{fig:fidelity}(a), the infidelity of this protocol (Control B) for the unitary dynamics smoothly decreases as the duration $T$ increases from zero, before saturating at zero for $T>T_0$. Applying this same protocol in the open system gives a similar overall behavior, with the key difference being that the saturated value of the infidelity at large $T$ is now non-zero. This behavior differs from that obtained for this control in a Markovian model with a constant decoherence rate, where slower processes produce higher infidelity, i.e. growing with $T$. In the ``Control B'' protocol, we have $h_z(t)=h_z$ and $h_x=0$, so that the model is the exactly solvable independent-boson model~\cite{mahan_many-particle_2013,viola_dynamical_1999}, which has non-Markovian decoherence. The constant infidelity for $T>T_0$ comes from the decoherence function of the independent-boson model, which approaches a non-zero constant for times $T\gg \tau_b\sim 1/\omega_c$.

Figure~\ref{fig:fidelity} shows that the optimization increases the performance for all process durations, as one would expect, but the increase becomes marked when $T>T_0$. This is because the fidelity in this regime is limited not by the distance to the target state and the speed of unitary evolution under $H_S$, but by the decoherence and dissipation produced by the bath. The improved fidelity thus corresponds to suppressing decoherence. 

An approach to suppressing decoherence one might consider is Dynamical Decoupling~\cite{viola_dynamical_1999} (DD). Such a protocol uses control fields to produce dynamics for the system that averages away the effects of the bath, which requires that the timescale of the system is much shorter than that of the bath $\tau_b$. This regime is not accessible in our optimization due to the bounds on the control fields. The standard DD sequence for the process we consider would be a train of $\pi$ pulses separated by a time $\tau\ll\tau_b$ and each with duration $\tau_p\ll\tau$; this would produce rapid spin-flips and so average away $H_{SB}\propto s_z$. Thus, standard DD requires a field $|h_{x}|\gg\pi/\tau_b$, which with $\tau_b\sim1/\omega_c\sim 0.3\,\mathrm{ps}$ is greater than $ h_x^{\mathrm{max}}$. 

To explore the mechanism that is giving the improvement in fidelity we compute a measure of non-Markovianity. There are many such interrelated measures~\cite{pollock_operational_2018,de_vega_dynamics_2017,li_concepts_2018} including, among others, those based on divisiblity and complete-positivity, and those based on information backflow. We choose that introduced by Breuer, Lane and Piilo~\cite{laine_measure_2010}, formed from the trace distance between pairs of states $D_{12}=\Tr |\rho_1(t)-\rho_2(t)|/2$, where $\rho_{1}(t)$ and $\rho_{2}(t)$ are obtained by time-evolving initial states $\rho_{1}$ and $\rho_{2}$ respectively. The non-Markovianity is \begin{equation}\mathcal{N}=\max_{\rho_1,\rho_2} \int \frac{dD_{12}}{dt} dt,\label{eq:nmmeasure}\end{equation} where the integral extends over regions where  the integrand is positive, and the maximum is over all pairs of orthogonal states on the surface of the Bloch sphere (which is the same as the maximum over all pairs of initial states, see \cite{wismann_optimal_2012}). 
The two curves in Fig.~\ref{fig:fidelity}(b) show the non-Markovianity as a function of process duration for the optimal controls (solid) and the constant field (dashed). For the constant field case, control B, the non-Markovianity increases from zero as the process duration increases from zero, becoming constant for $T\gtrapprox 2\,\mathrm{ps}$. This corresponds to the build-up of correlations due to the system-bath coupling in the independent boson model, and is unaffected by the system dynamics since $[H_S,H_{SB}]=0$ when $h_x=0$. The non-Markovianity with the optimal control A is higher, but initially shows a similar increase and saturation as the duration $T$ increases from zero. However, there is then a marked feature at $T\approx T_0$, beyond which the non-Markovianity increases rapidly with process duration before reaching a new and much higher level. This suggests that the improvement in fidelity for long process durations occurs because the optimization increases the degree of non-Markovianity of the map, restoring some of the information lost to the environment during the earlier parts of the dynamics by the end of the process. 

\begin{figure}
\includegraphics{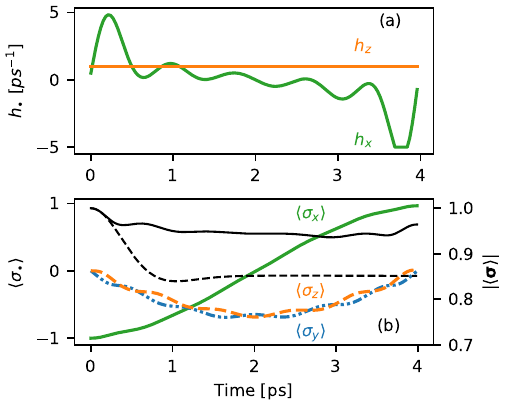}
\caption{\label{fig:sampledynamics} (a) Driving fields $h_x(t)$ (green) and $h_z(t)$ (orange) for the optimal state transfer process, control A, of duration 4.0 ps. (b) Dynamics of the components of the Bloch vector (left axis, green/solid: $\expval{\sigma_x}$, blue/dot-dashed: $\expval{\sigma_y}$, orange/dashed: $\expval{\sigma_z}$), and length of the Bloch vector (right axis, black/solid). The length of the Bloch vector for control B
is also shown (right axis, black/dashed).}
\end{figure}

Figure~\ref{fig:sampledynamics} shows the optimized control fields (top panel), and the dynamics of the Bloch vector $\langle\sigma_{x,y,z}\rangle$, for the optimal protocol, control A, at $T=4.0\,\mathrm{ps}$. For this duration, the optimal $h_z$ remains constant, while the optimized $h_x$ has an overall slope and oscillations of varying magnitude. The resulting dynamics of the Bloch vector moves down below the equator during the control and develops weak oscillations. The clearest effect of the optimization is seen in the length of the Bloch vector, which is shown for both protocols A and B. In the latter case (constant field) the length of the Bloch vector is given by the decoherence function of the independent boson model, which goes from one, for zero time, to a constant non-zero value at late times. The optimal control A clearly avoids this decay and even shows an increase in the Bloch-vector length at the very end of the pulse, consistent with a decoherence suppression mechanism. The control in $h_x$ is very different from that for a standard DD sequence as described above. The behavior shown is representative of that we find for other process durations~\cite{supplement}. For a unitary dynamics it has been shown that there is a unique global optimum for durations near the quantum speed limit~\cite{larocca_quantum_2018}. The consistency in the form of our solutions suggests that this may also be the case here. 

Some insight into a possible mechanism for the improved fidelity can be gained by considering the decoherence at zero temperature when $h_x=h_z=0$. This is the solvable independent-boson model~\cite{mahan_many-particle_2013,viola_dynamical_1999}, in which an initial product state $\ket{\uparrow_x}\otimes|0\rangle=(\ket{\uparrow}+\ket{\downarrow})\otimes\ket{0}/\sqrt{2}$ evolves to $(\ket{\uparrow}\otimes\ket{f}+\ket{\downarrow}\otimes\ket{-f})/\sqrt{2}$. Here $|\pm f\rangle$ are bath states in which the Gaussian ground-state wavefunctions of the oscillators have evolved in oppositely-displaced potentials. This reduces the system coherence by the overlap of the bath states, $\langle \sigma_x\rangle=2\Re \braket{f}{-f}$. We suggest that our controls both suppress and reverse this polaron formation process, disentangling the system and environment. For a single-oscillator environment such disentangling occurs periodically without the control fields; our optimization may be exploiting similar physics in an ensemble. Optimization results for other spectral densities and temperatures~\cite{supplement} are consistent with this suggestion.


In conclusion, we have presented a method for computing optimal controls in general non-Markovian systems. It extends the PT method~\cite{fux_efficient_2021} to efficiently compute gradients, so permitting optimization over large numbers of control parameters. The method takes full account of the changes in the dissipator produced by the control fields, allowing it to discover protocols which exploit non-Markovianity for improved performance. We have used it to investigate state transfer in the spin-boson model, where we find that slow processes can improve on fast ones by exploiting non-Markovian effects. These results show that performance improvements are possible across the many qubit implementations subject to non-Markovian decoherence, and could be identified using the methods described here.

We acknowledge discussions with F. Binder.
E.\,P.\,B. acknowledges support from the Irish Research Council (GOIPG/2019/1871).  G.\,E.\,F. acknowledges support from EPSRC (EP/L015110/1). B.\,W.\,L. and J.\,K. acknowledge support from EPSRC (EP/T014032/1). P.\,R.\,E. acknowledges support from Science Foundation Ireland (21/FFP-P/10142). In order to meet institutional and research funder open access requirements, any accepted manuscript arising shall be open access under a Creative Commons Attribution (CC BY) reuse licence with zero embargo.

\nocite{campaioli_tight_2019}

%

\end{document}



\title{Supplemental Material for ``Optimizing performance of quantum operations with non-Markovian decoherence: the tortoise or the hare?''}

\title{Optimizing performance of quantum operations with non-Markovian decoherence: \\ the tortoise or the hare?}
\author{Eoin P. Butler}
 \affiliation{School of Physics, Trinity College Dublin, College Green, Dublin 2, Ireland}
 \affiliation{Trinity Quantum Alliance, Unit 16, Trinity Technology and Enterprise Centre, Pearse Street, Dublin 2, Ireland}
\author{Gerald E. Fux}%
\affiliation{SUPA, School of Physics and Astronomy, University of St Andrews, St Andrews KY16 9SS, United Kingdom}%
\affiliation{Abdus Salam International Center for Theoretical Physics (ICTP), Strada Costiera 11, 34151 Trieste, Italy}%
\author{Carlos Ortega-Taberner}
 \affiliation{School of Physics, Trinity College Dublin, College Green, Dublin 2, Ireland}
 \affiliation{Trinity Quantum Alliance, Unit 16, Trinity Technology and Enterprise Centre, Pearse Street, Dublin 2, Ireland}
\author{Brendon W. Lovett}
\affiliation{SUPA, School of Physics and Astronomy, University of St Andrews, St Andrews KY16 9SS, United Kingdom}
\author{Jonathan Keeling}
\affiliation{SUPA, School of Physics and Astronomy, University of St Andrews, St Andrews KY16 9SS, United Kingdom}%
\author{Paul R. Eastham}
 \affiliation{School of Physics, Trinity College Dublin, College Green, Dublin 2, Ireland}
 \affiliation{Trinity Quantum Alliance, Unit 16, Trinity Technology and Enterprise Centre, Pearse Street, Dublin 2, Ireland}

\date{\today}
\maketitle

This supplemental material is organized as follows. In Sec.\ \ref{sec:suppintro} we provide a brief general introduction to the process tensor formalism and some additional details of the method. It also includes details of the computational cost of the simulations. Section~\ref{sec:suppgrad} provides additional details of the computation of the gradient of an objective function. Section~\ref{sec:suppqsl} presents some results on quantum speed limits. Section~\ref{sec:suppresults} gives further supplementary results, including optimal controls and dynamics for different spectral densities. It also describes the additional supplementary results, for a range of process durations, that are provided in a separate file.

\section{Brief introduction to process tensors}
\label{sec:suppintro}

\begin{figure}
    \centering
    \includegraphics{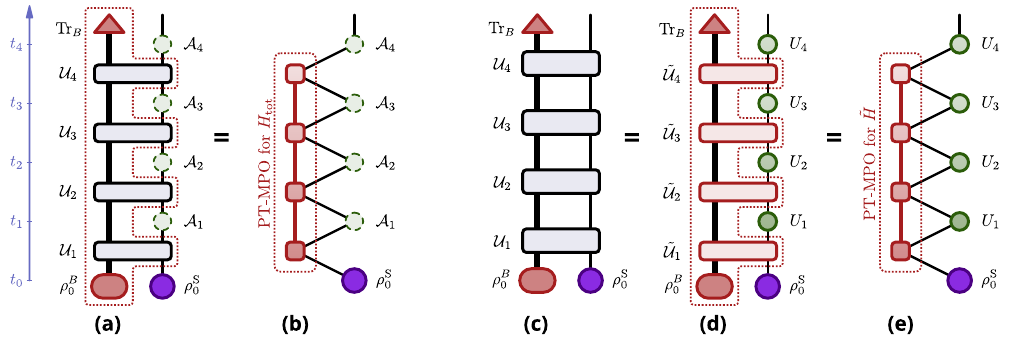}
    \caption{\label{fig:process-tensor}%
    Tensor network diagrams representing the reduced system state $\rho^\mathrm{S}(t_4)$ at time $t_4$.
    Panels (a) and (b) show $\rho^\mathrm{S}(t_4)$ after the application of a control sequence $\{\mathcal{A}_1, \ldots, \mathcal{A}_4 \}$.
    Panels (c), (d), and (e) show $\rho^\mathrm{S}(t_4)$ after an evolution that can be separated into brief unitary propagators $\tilde{\mathcal{U}}_n$ due to the environment interaction $\tilde{H} = H_{SB} + H_B$ and brief unitary propagators $U_n$ due to the pure system part $H_\mathrm{S}(t)$.
    The red dotted regions show the process tensors in matrix product operator form (PT-MPO).
    The red triangle marked $\mathrm{Tr}_B$ denotes the partial trace over the environment, and $\rho^B_0$ and $\rho^\mathrm{S}_0$ denote the initial environment and system states, respectively.}

\end{figure}

The process tensor formalism is an operational approach to general (i.e. not restricted to Markovian) open quantum systems~\cite{pollock_non-markovian_2018}.
The key idea is to describe open quantum system through a mathematical object---the process tensor (PT)---that encodes the outcome of all possible experiments that consist of a sequence of \emph{control operations} on the system.
A control operation may be any experimentally realizable intervention on the system, i.e. a completely positive map, which for example includes measurements on the system, or the application of a unitary.

In Fig.~\ref{fig:process-tensor}a we draw a quantum circuit in Liouville space for a general total system (i.e. the open system together with its environment) with control operations on only the open system at times $t_1$, $t_2$, $t_3$, and $t_4$.
The super-operators $\mathcal{U}_n=\exp\left[\int_{t_{n-1}}^{t_{n}} \mathcal{L}_\mathrm{tot}(t) \mathrm{d}t \right]$ are the propagators between time $t_{n-1}$ and $t_n$, where $\mathcal{L}_\mathrm{tot}(t) \: \cdot = - i \left[ H_\mathrm{tot}(t) , \:\cdot\:\right]$ is the Liouvillian for the total system.
Because we assume that the experimenter has no direct access to the environment we trace it out at the final time.
The object that encodes the outcome of all different control sequences $\{\mathcal{A}_1, \mathcal{A}_2, \mathcal{A}_3, \mathcal{A}_4\}$ can be easily identified as the red dotted region in Fig.~\ref{fig:process-tensor}a.
This is the process tensor as introduced in~\cite{pollock_non-markovian_2018}.
Through a reinterpretation of this quantum circuit as a tensor network one may note that the process tensor naturally assumes the form of a matrix product operator (PT-MPO).
Note that the particular MPO representation in Fig.~\ref{fig:process-tensor}a uses the environment Liouville space as bonds, which is prohibitively large for most systems of interest.
However, because there is no need to explicitly keep track of the environment, there often exists an efficient representation of the PT-MPO with low bond dimensions, which we draw in Fig.~\ref{fig:process-tensor}b.
As mentioned in the main text, there are several numerical methods that allow such a PT-MPO for multiple different types of environments to be obtained~\cite{makri_tensor_1995, makri_tensor_1995-1, banuls_matrix_2009, jorgensen_exploiting_2019, strathearn_modelling_2020, lerose_influence_2020, fux_efficient_2021, cygorek_simulation_2022, sonner_influence_2021, ye_constructing_2021, white_non-markovian_2021, thoenniss_non-equilibrium_2022, OQuPy2022, thoenniss_efficient_2022, ng_real_2022}.
The necessary bond dimensions of an PT-MPO is related to the degree of non-Markovianity of the interaction between the system and environment~\cite{pollock_operational_2018}.

In this paper we make use of a Suzuki-Trotter expansion~\cite{trotter_product_1959} of the unitary evolution of the total system \mbox{$H_\mathrm{tot} = H_\mathrm{S}(t) + H_B + H_{SB}$} into an environment interaction part and a pure system part. For small enough time steps the unitary evolution $\mathcal{U}_n$ can be separated into the unitary evolution $\tilde{\mathcal{U}}_n$ due to the environment interaction $\tilde{H} = H_{SB} + H_B$ and the unitary evolution $U_n$ due to the pure system part $H_\mathrm{S}(t)$ as shown in Fig.~\ref{fig:process-tensor}c~and~\ref{fig:process-tensor}d.
One can then view the pure system propagators $U_n$ as control operations and the rest as a process tensor for a total system with the reduced total Hamiltonian $\tilde{H} = H_{SB} + H_B$ as shown in Fig.~\ref{fig:process-tensor}d.

For the computations in the main text we employed the method introduced in~\cite{jorgensen_exploiting_2019} to obtain such a PT-MPO representation of the Trotterized time evolution. We used the implementation in the OQuPy package~\cite{OQuPy2022} with a time step $\Delta t = 0.03\ \text{ps}$, a memory cut-off of $K=60$ timesteps, and a relative singular value decomposition cutoff threshold of~$10^{-7}$. The most demanding calculations in the text are those for the longest duration, $T=5.0\;\mathrm{ps}$. The computation of the process tensor for this case took approximately $7$ minutes on a laptop (Macbook Pro with an 8 core Intel i9-9880H processor), and produced a PT-MPO with a maximal bond dimension of 234 which occupied 1 GB of memory. Propagating a state using this PT took 25 seconds. The optimization, over $N=332$ parameters, took approximately 45 minutes and converged in 37 iterations. The shorter durations are significantly easier: for $T=1.0\;\mathrm{ps}$ computing the process tensor took less than 1 second. The optimization, over $N=66$ parameters, took less than 3 seconds and converged in 6 iterations.

We emphasize that the method introduced in this work is not restricted to the particular choice of the environment or the particular choice of algorithm used to compute the PT-MPO.
The method only relies on the availability of the environment's PT-MPO with a moderate bond dimension, which may be obtained for a variety of different environments through a variety of different methods.

\section{Construction of the gradient from the process tensor}

\label{sec:suppgrad}

In this section we present details of how we calculate the gradient of an objective function from the PT-MPO. To establish notation we first recall the calculation of the time evolution from an initial to a final state, shown in Fig.~\ref{fig:process-tensor}e and Fig. 1a of the main text. Recall that states here are vectorized system density matrices. The time evolution proceeds in discrete time steps at times $t_n=n\Delta t$ for $n=1,\ldots,N_t$. For a system with Hilbert space dimensionality $d_{H_S}$ the state at time $t_n$ is a $d_{H_S}^2$ component vector which we denote $\rho^i_n$ ($i=1,\ldots d_{H_S}^2$). The system propagator over the $n^{th}$ time step is a $(d_{H_S}^2,d_{H_S}^2)$ matrix we denote $U^{ij}_n$. 

Fig~\ref{fig:process-tensor}e shows the evolution over four time steps with time increasing from bottom to top. The state at the final time can be constructed by evaluating the contractions in this network forwards in time. One first contracts the lowest external leg of the PT-MPO with the leg of the initial state. The next two external legs of the PT-MPO are, in order, an output from it and an input to it. The output (input) leg of the PT-MPO is joined to the input (output) leg of the system propagator for the first time step. The propagators for the the next two time steps are included in a similar manner, thus interleaving the factors in the path integral due to the environment and the system Hamiltonian. The final step is a contraction of the last output leg of the PT-MPO with the input leg of the last system propagator. The resulting diagram has one open leg, the output leg of the last system propagator, and represents the final state.

In the main text we considered a particular objective function, the fidelity $\mathcal{F}$, and took as control parameters the fields $h_{x,z}(t_n)$. Here we consider a more general case where the objective is a function $Z(\rho^i_f=\rho_{N_t}^i)$ of the final state, and there are $N$ control parameters $c_a$. In general each propagator $U^{ij}_n$ can depend on all the control parameters. Thus we have \begin{equation}\label{eq:chainrule}
    \frac{\partial Z}{\partial c_a} = \sum_{n} \sum_{i,j,k}^{d_{H_S}^2} 
    \frac{\partial Z}{\partial \rho_f^i} \frac{\partial\rho_f^i}{\partial U^{jk}_n} \frac{\partial U^{jk}_n}{\partial c_a}.\end{equation} 
    
\begin{figure}
    \centering
    \includegraphics{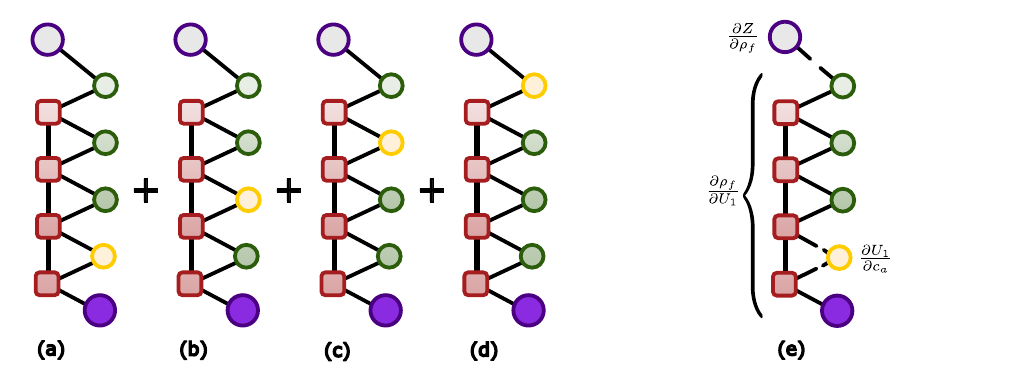}
    \caption{Diagrammatic representation of the derivative of an objective function with respect to a control parameter, Eq. (\ref{eq:chainrule}). (a) is the contribution involving the propagator over the first time step, $U_1$, and (b--d) the contributions from subsequent steps. The yellow circle in each panel is the derivative of the propagator over that time step with respect to the control parameter. The purple circle at the top of each panel is the derivative of the objective function with respect to the final state. (e) shows how (a) is constructed by joining legs (contracting indices) among tensors corresponding to the factors in Eq. (\ref{eq:chainrule}).
    \label{fig:chainrule}}
\end{figure}

Fig.\ \ref{fig:chainrule} shows Eq. (\ref{eq:chainrule}) for four time steps. The terms involving the propagators $U_1,\ldots,U_4$ are shown in panels (a)--(d) respectively. Panel (e) shows how panel (a) is formed from the three factors in Eq. (\ref{eq:chainrule}). The rank-1 tensor $Z^i\equiv\frac{\partial Z}{\partial \rho^i_f}$ is the purple circle at the top of the diagram, and the rank-2 tensor $\frac{\partial U^{jk}_1}{\partial c_a}$, for a given $a$, is the yellow circle at the first time step. The remainder of the diagram is the rank-3 tensor $\frac{\partial \rho_f^i}{\partial U_1^{jk}}$. As discussed in the main text, the PT is a multilinear map from the propagators to the final state. Thus the diagram for $\frac{\partial \rho_f^i}{\partial U_1^{jk}}$, for example, is the same as that for the final state, with the first propagator, $U_1$, omitted. Furthermore, the contraction of this tensor with $Z^i$ is the derivative of the objective function with respect to the first propagator, $\frac{\partial Z}{\partial U_1^{jk}}$. This important object is the diagram in Fig.\ \ref{fig:chainrule}a with the yellow circle omitted. The derivatives with respect to the propagators at other time steps are obtained similarly from Figs.\ \ref{fig:chainrule}b--d. 

\begin{figure}
\includegraphics[width=\linewidth]{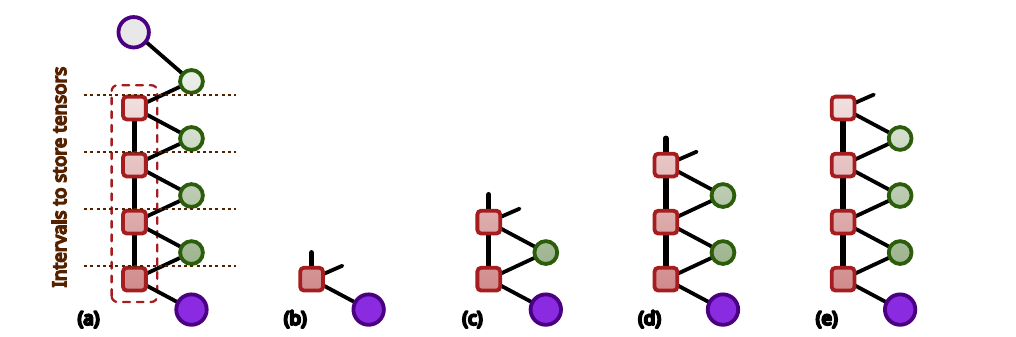}
\caption{Illustration of the forward propagation calculation for a process tensor with four time steps. The left-hand diagram is the scalar product of the time-evolved state with the derivative of the objective function with respect to that state (top purple circle). In forward propagation the initial state is evolved forwards in time, to successively compute the diagrams shown in panels (b--e). To compute the gradient these tensors are stored.\label{fig:forwardprop}}
\end{figure}

\begin{figure}
\includegraphics[width=\linewidth]{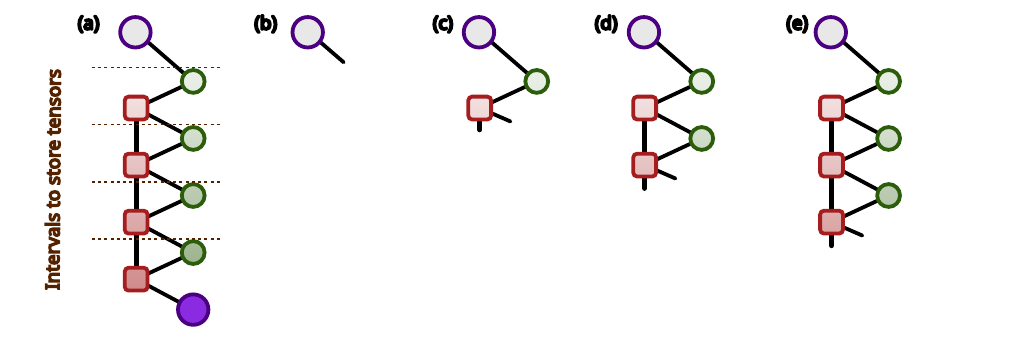}
\caption{\label{fig:backprop}  Illustration of the back propagation calculation for a process tensor with four time steps. The left-hand diagram is the scalar product of the time-evolved state with the derivative of the objective function with respect to that state (top purple circle). In back propagation this tensor is evolved backwards in time, to successively compute the diagrams shown in panels (b--e). The gradient is obtained by combining these tensors with those stored during the forward propagation (see text).} 
\end{figure}

To calculate Eq. (\ref{eq:chainrule}) we first compute the derivatives of the objective function with respect to the system propagators. This is done using two computations of the tensor network: a forward propagation, shown in Fig.~\ref{fig:forwardprop}, and a back propagation, shown in Fig.~\ref{fig:backprop}. In the forward propagation, we evaluate the network starting from the initial state and proceeding forward in time, as is done to compute the objective function itself. The only change is that we store the tensors obtained at each time step, shown in Figs.\ \ref{fig:forwardprop}b--e. Each stored tensor has two legs, one internal leg within the PT-MPO, known as a `bond' leg, and one external leg to the PT-MPO, which would normally have been connected to a system propagator, known as a `state' leg. The back propagation is shown in Fig.\ \ref{fig:backprop}. One evaluates this diagram starting from the top with the tensor $Z^i$ and proceeding backwards in time. One again retains the intermediate forms computed at each time step, shown in Figs.\ \ref{fig:backprop}b--e. We see that by joining the bond leg in Fig.\ \ref{fig:forwardprop}b with that in Fig.\ \ref{fig:backprop}e we obtain the derivative $\frac{\partial Z}{\partial U_1}$.  The other derivatives are obtained similarly from pairs of tensors stored during the forward and back propagation. These derivatives
are then combined with the factors $\frac{\partial U_n}{\partial c_a}$ to give the gradient. 

To connect these considerations to the special case considered in the main text, we note that there we wish to maximize the fidelity, $\mathcal{F}=(\mathrm{Tr} \sqrt{\sqrt{\rho_f}\sigma\sqrt{\rho_f}})^2$, between the density matrices describing the target state, $\sigma$, and the final system state, $\rho_f$. For a pure target state the density matrix is $\sigma=\ketbra{\sigma}{\sigma}$, and the fidelity simplifies to $\mathcal{F}=\langle\sigma|\rho_f|\sigma\rangle$. In Liouville space this is the scalar product of the vectorized forms of the final density matrix and the transposed target density matrix, $\mathcal{F}=(\sigma^T)^i \rho^i_f$. Thus, $Z^i=(\sigma^T)^i$, and the gradient of the fidelity is obtained by back propagation from the transposed target state as claimed. An additional simplification can be made when each control parameter only affects the propagator of a single time step (as in the presented example in the main text). In this case, for each of the control parameters only the corresponding time step $n$ in Eq.\ \ref{eq:chainrule} will contribute to the derivative of the objective function.

\section{Quantum speed limits}

\label{sec:suppqsl}

\begin{figure}
    \centering
    \includegraphics[width=0.5\linewidth]{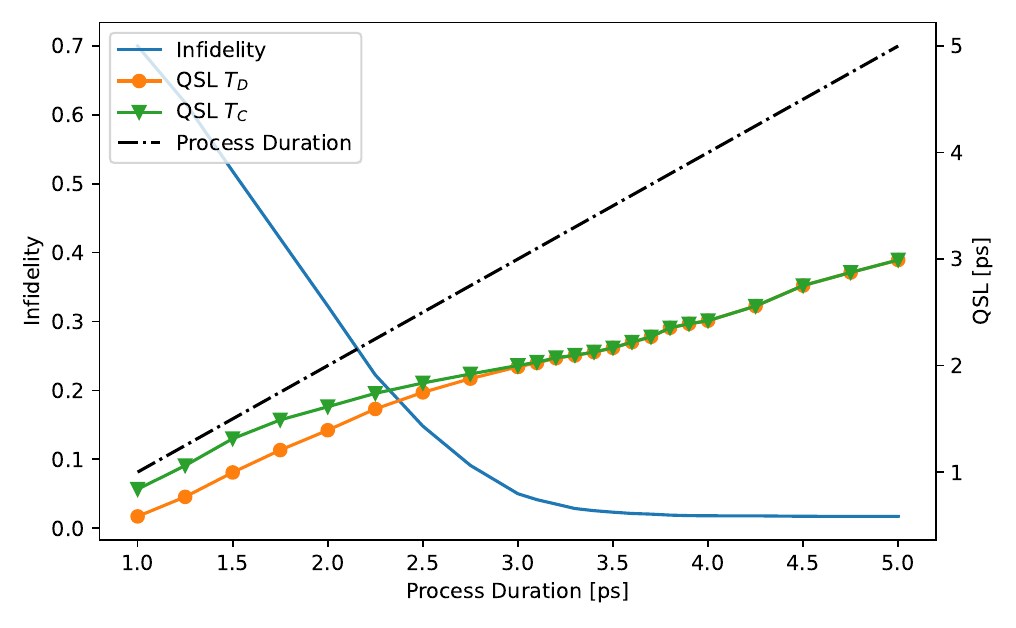}
    \caption{Quantum speed limits (right axis) for the optimal protocol, control A, as a function of process duration. Green/triangles: Campaioli et al. speed limit. Orange/circles: Deffner et al. speed limit. Black dashed line: process duration. The solid blue line shows, on the left axis, the infidelity.\label{sfig:qsl}}
    \label{fig:enter-label}
\end{figure}

Quantum speed limits (QSLs) are upper bounds on the speed of evolution between one quantum state and another, and therefore lower bounds on the time required for such a process. For unitary dynamics the unified Margolus-Levitin and Mandelstamm-Tamm bound is tight for certain states and, as discussed in the main text, gives the minimum time for the state transfer in the case of a closed system. Although various bounds have been obtained for open quantum systems, they are not expected to be tight for general evolutions~\cite{campaioli_tight_2019}. Nonetheless, it is interesting to consider the relation between the time of our optimal process and some QSLs for open systems. Fig.~\ref{sfig:qsl} shows the QSLs of Campaioli et al.~\cite{campaioli_tight_2019}, $T_{C}$, and Deffner et al.~\cite{deffner_quantum_2013}, $T_{D}$, for the evolutions obtained by our optimization, as a function of the process duration $T$. We also show the fidelity, as in the main text.  In the region of short process durations the tightest of the two bounds, $T_{C}$, reaches a significant fraction (around 80\%) of the process duration, but the limit is not reached. As we increase the process duration, the bounds increase more slowly than the process duration, and the state transfer is operating further and further from the open-system QSL. This is in line with our conclusion in the main text, that non-Markovian effects can lead to a trade-off between speed and fidelity.

\section{Supplemental results}
\label{sec:suppresults}

\subsection{Dynamics of information backflow}

\begin{figure}[h]
  \centering
  \includegraphics[width=0.5\linewidth]{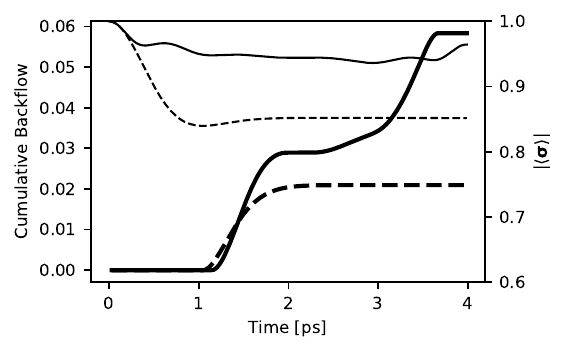}
  \caption{Cumulative information backflow as a function of time (left axis, thick lines) and Bloch vector lengths (right axis, thin lines), for the optimized control (control A, solid lines) and the unoptimized one (control B, dashed lines). Results are shown for the state transfer process of duration 4 ps.  }
  \label{fig:supp-backflow}
\end{figure}

In Fig.~\ref{fig:supp-backflow}, we explore how the information backflow evolves during the controls. The measure of non-Markovianity we use is defined by Eq. (1) in the main text; recall it is the cumulative increase in trace distance between trajectories starting from two states, calculated over the entire process, and maximized over all pairs of starting states. To explore how the information backflow varies with time for a given dynamics, we consider the pair of trajectories which maximize this measure over the entire process, and plot the cumulative increase in trace-distance up to a given time. The thick solid and dashed curves in Fig.~\ref{fig:supp-backflow} show this measure for the optimized control A and the unoptimized control B, as in the main text, for a process duration 4 ps. We also show, for comparison, the length of the Bloch vector under the two dynamics. It can be seen that, in the optimized process, information backflow is present for most of the process. This is excepting a small part at the beginning, when there would be in any case no information lost to the environment to recover, and at the end, when presumably all the information which could be restored from the environment has been.

\subsection{Optimization for alternate spectral densities}

\begin{figure}[ht]
\includegraphics{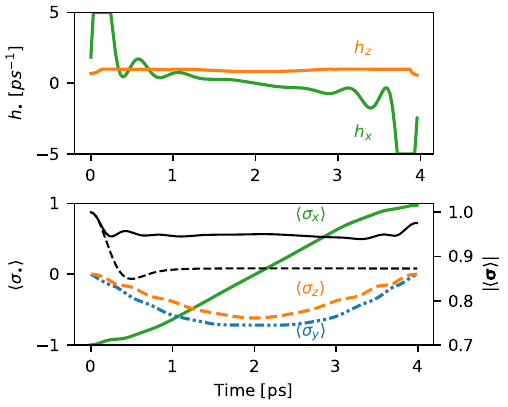}
\includegraphics{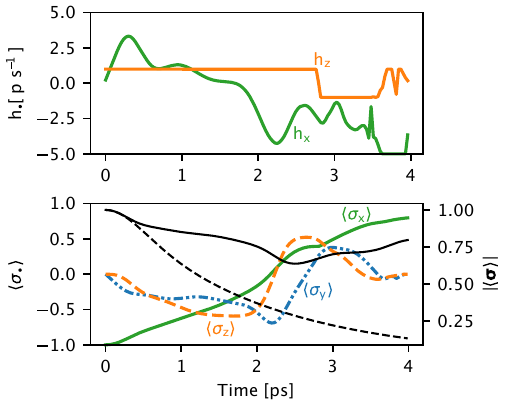}
\caption{Optimized pulses and dynamics for different spectral densities. Both plots correspond to Fig. 3 of the main text, with a changed spectral density. For those on the left the cut-off is doubled to $6.08\,\text{ps}^{-1}$. For those on the right the super-Ohmic spectral density is changed to the corresponding Ohmic form, $J(\omega)=2\alpha\omega\exp\left[-\omega^2/\omega_c^2\right]$. The top panels show the driving fields $h_x(t)$ (green) and $h_z(t)$ (orange) for the optimal state transfer process, control A, of duration 4.0 ps. The lower panels show the dynamics of the components of the Bloch vector (left axis, green/solid: $\expval{\sigma_x}$ , blue/dot-dashed: $\expval{\sigma_y}$, orange/dashed: $\expval{\sigma_z}$), and length of the Bloch vector (right axis, black/solid). The length of the Bloch vector for control B is also shown (right axis, black/dashed).\label{supp:sdvaryfig}}
\end{figure}
\begin{figure}
\includegraphics[width=0.5\linewidth]{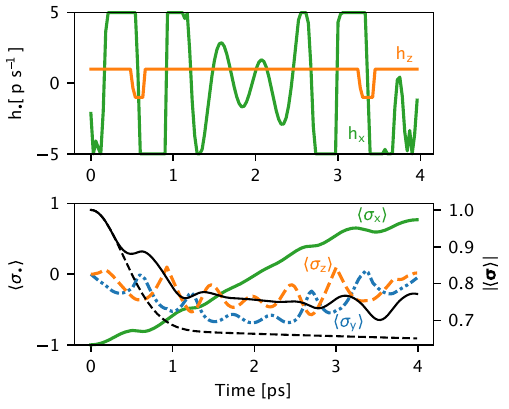}
\caption{Optimized pulse and dynamics at a temperature of 20 K. The plots correspond to Fig. 3 of the main text, and the parameters other than temperature are the same.\label{supp:hightfig}}
\end{figure}

In Figs.~\ref{supp:sdvaryfig} and \ref{supp:hightfig} we show how the optimal controls, and resulting dynamics, are affected by changing some aspects of the environment. These plots correspond to Fig.~3 of the main text, with different environment parameters. As discussed in the main text, we suggest that the optimized protocols act to partially undo the formation of the polaron, which occurs as different bath modes go out of phase with one another. This suggests that the oscillations in the control visible in Fig.~3 of the main text may correspond to the high-frequency cut-off of the spectral density,  $\omega_c$. The left plots in Fig.~\ref{supp:sdvaryfig} show the results of the optimization with twice the cut-off frequency, and it can be seen that the frequency of the oscillations in the control has indeed doubled. The right plots show results for an Ohmic spectral density (with the original cut-off), where we see that the optimal control is noticeably slower than the super-Ohmic case. This is consistent with the more significant role of low-frequency modes in Ohmic baths generally. The improvement produced by the optimization here is remarkable, given it otherwise has much stronger decoherence, as can be seen in the lengths of the Bloch vectors for the two cases. Finally, Fig.~\ref{supp:hightfig} shows the results at a higher temperature. This leads to stronger controls, and a lower final fidelity, than are obtained at lower temperatures. However, there is still a significant improvement after optimizing for the non-Markovian case.

\subsection{Description of supplementary results}

Fig. 3 of the main text showed the optimal controls and dynamics for a process duration of 4 ps. The file \verb|supplementary_results.pdf| shows the corresponding results for a range of process durations from 1 to 5 ps, with each page corresponding to a different duration. One can view this file in single-page mode and move between pages to generate an animation. The bottom left panel, which is repeated on every page, reproduces Fig. 2a of the main text. The infidelity of the optimal control (control A) is shown as a function of process duration as the blue curve. On each page one point on that curve is marked with a red cross. This indicates the process duration for which results are shown in the remaining panels. These are the computed optimal controls (top left), the time-dependence of the expectation values (top right) and the trajectory in the Bloch sphere (bottom right). The top right panel also shows the length of the Bloch vector for the unoptimized control (control B).

\newsavebox\mytempbib
\savebox\mytempbib{\parbox{\textwidth}{
}}